\documentclass{pnastwo}

\usepackage[dvips]{graphicx}

\usepackage{amssymb,amsfonts,amsmath}

\contributor{Submitted to Proceedings
of the National Academy of Sciences of the United States of America}
\url{www.pnas.org/cgi/doi/10.1073/pnas.0709640104}
\copyrightyear{2008}
\issuedate{Issue Date}
\volume{Volume}
\issuenumber{Issue Number}

\begin{document}



\title{Supermassive Black-Hole Growth Over Cosmic Time: Active Galaxy
Demography, Physics, and Ecology from Chandra Surveys}





\author{W.N. Brandt\affil{1}{Department of Astronomy \& Astrophysics, 
The Pennsylvania State University, 525 Davey Lab, University Park, PA 16802, USA}\and
D.M. Alexander\affil{2}{Department of Physics, Durham University, 
South Road, Durham DH1 3LE, UK}}

\contributor{Submitted to Proceedings of the National Academy of Sciences
of the United States of America}

\maketitle

\def\simgt{\lower 2pt \hbox{$\, \buildrel {\scriptstyle >}\over {\scriptstyle \sim}\,$}}
\def\simlt{\lower 2pt \hbox{$\, \buildrel {\scriptstyle <}\over {\scriptstyle \sim}\,$}}

\def\astroh{{\it Astro-H\/}}
\def\chandra{{\it Chandra\/}}
\def\chandras{{\it Chandra's\/}}
\def\erosita{{\it eROSITA\/}}
\def\exist{{\it EXIST\/}}
\def\herschel{{\it {\it Herschel}\/}}
\def\hst{{\it {\it HST}\/}}
\def\ixo{{\it IXO\/}}
\def\jwst{{\it JWST\/}}
\def\nustar{{\it NuSTAR\/}}
\def\rosat{{\it ROSAT\/}}
\def\spitzer{{\it Spitzer\/}}
\def\wfxt{{\it WFXT\/}}
\def\xmm{{\it XMM-Newton\/}}

\def\todo{{\Huge $\bullet$}}
\def\aox{$\alpha_{\rm ox}$}

\begin{article}

\begin{abstract} 
Extragalactic \hbox{X-ray} surveys over the past decade have 
dramatically improved understanding of the majority populations of 
active galactic nuclei (AGNs) over most of the history of the 
Universe. Here we briefly highlight some of the exciting 
discoveries about AGN demography, physics, and ecology with a 
focus on results from \chandra. We also discuss some key 
unresolved questions and future prospects. 
\end{abstract}







\section{Relevant Capabilities of X-ray Surveys and Chandra}

Extragalactic \hbox{X-ray} surveys are powerful for studying the growing 
supermassive black holes (SMBHs) in AGNs for several reasons. 
First, \hbox{X-ray} emission is empirically found to be nearly universal from 
luminous AGNs; the accretion disk and its ``corona'' are robust even if their
details remain somewhat mysterious. 
Second, \hbox{X-ray} emission is penetrating and has reduced absorption bias 
compared to, e.g., optical and UV emission. This is critically important
since it is now known that the majority of AGNs suffer from significant
intrinsic obscuration. Furthermore, the level of \hbox{X-ray} absorption bias
drops toward high redshift, since increasingly penetrating rest-frame 
\hbox{X-rays} are observed.  
Finally, \hbox{X-ray} observations maximize the contrast between SMBH-related 
light and host-galaxy starlight. Having such high contrast is crucial
when studying high-redshift objects that cannot be resolved
spatially. \hbox{X-ray} surveys thus provide the ``purest'' AGN samples; 
most ($\simgt 80$\%) of the sources even in the deepest \hbox{X-ray} 
observations are AGNs, while few ($\simlt 10$\%) of the sources in the 
deepest optical and infrared observations are AGNs. 

The unmatched angular resolution, low background, broad bandpass, 
and respectable field of view of \chandra\ have provided dramatic 
advances in our ability to survey the \hbox{X-ray} emission from AGNs over 
most of the history of the Universe. 
The deepest \chandra\ observations are \hbox{50--250} times more
sensitive than those of previous missions
(the exact factor depending upon the bandpass considered), 
detecting sources with photon fluxes as low as one count per 5~days.
Source positions measured by \chandra\ are typically reliable to
within \hbox{0.2--0.5$^{\prime\prime}$}; 
this is essential for robust identifications and follow-up work 
at faint fluxes. 
The surveys executed by \chandra\ have each detected hundreds-to-thousands 
of sources, allowing statistically meaningful studies of source populations. 
Finally, the well-maintained data archive allows the effective
federation of \chandra\ surveys by any astronomer to address 
specific scientific questions of interest. 

\begin{figure}[t!]
\centerline{\includegraphics[width=.4\textwidth,angle=-90]{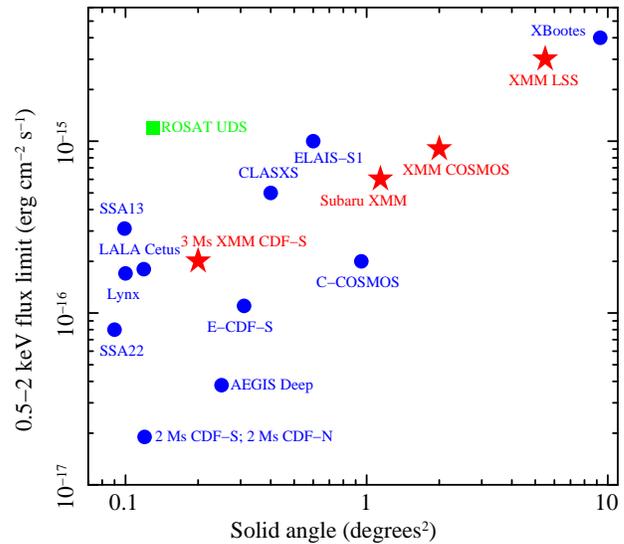}}
\caption{Distribution of some well-known extragalactic surveys 
by Chandra (blue dots), XMM-Newton (red stars), and ROSAT (green square) 
in the \hbox{0.5--2~keV} flux-limit versus solid-angle plane. 
Each of the surveys shown has a range of flux limits across its
solid angle; we have generally shown the most sensitive flux limit. 
All surveys shown are contiguous. 
Adapted from Brandt \& Hasinger (2005).}\label{fig1}
\end{figure}

Currently about 35 \chandra\ and \xmm\ surveys have been performed
that cover most of the practically accessible ``discovery space'' 
of sensitivity vs.\,solid-angle. These include contiguous surveys, 
many of which are shown in Fig.~1, as well as the equally important
non-contiguous and often serendipitous surveys (e.g., ChaMP, 
HELLAS2XMM, SEXSI, the \xmm\ SSC surveys). 
Enormous progress has been made over the past decade in obtaining 
identification spectra for large, representative samples of the 
detected sources; this work has often utilized the largest 
ground-based telescopes on Earth (e.g., Gemini, Keck, Subaru, 
VLT). However, spectroscopic identification remains a persistent 
challenge and bottleneck, especially at faint fluxes \hbox{($I=24$--28)}, 
and serves as one important driver for future 
Extremely Large Telescopes (ELTs). 
Multiwavelength observations of the \chandra\ survey sources, from 
the radio to the UV, have also been critical for advancing 
understanding, as expected given the broad-band nature of AGN 
emission. These have 
improved the reliability of the \hbox{X-ray} source identifications,
allowed the derivation of high-quality photometric redshifts, 
constrained AGN accretion physics, 
measured host-galaxy properties, 
assessed the relative importance of SMBH vs.\,stellar power, and
even discovered AGNs that were missed by
\chandra\ (e.g., due to extreme obscuration). 

Below we will briefly highlight some of the exciting discoveries from 
\chandra\ surveys about the tightly related topics of AGN demography, 
physics, and ecology. 
Our focus will be on \chandra\ results from the past decade, as befits 
this 10th birthday symposium for \chandra, implicitly also recognizing 
the fundamental advances made by extragalactic surveys with \xmm. 
Comparisons will sometimes be made with the community's understanding 
at around the time of the \chandra\ launch in mid-1999, since these 
illustrate just how dramatic the advances have been over
\chandras\ first decade of discovery. 
Furthermore, due to limited space, our references to the literature 
will necessarily be limited, highly selective, and incomplete; our 
humble apologies in advance.  


\section{Demography}

From the 1960's to the 1990's, the study of AGN evolution 
was dominated by wide-field optical surveys of rare, luminous quasars
(e.g., Boyle 2001; Osmer 2004). These were 
found to peak in comoving number density at \hbox{$z\approx 2$--3} 
and showed evolution consistent with pure luminosity evolution 
models. These surveys left open a major question: how does the 
{\it numerically dominant\/} population of moderate-luminosity AGNs 
evolve? Many astronomers expected, prior to the launch of \chandra, 
that moderate-luminosity AGNs would evolve in the same manner as
luminous quasars. 

However, even from the \rosat\ soft \hbox{X-ray} extragalactic 
surveys, hints were emerging that AGN evolution is 
significantly luminosity dependent (e.g., Miyaji et~al.\ 2000). These 
surveys also hinted, independently, that the \hbox{X-ray} selected quasar space 
density at $z\simgt 3$ might not decline in the manner seen for optically 
and radio selected quasars. The observational constraints, at the time of 
the \chandra\ launch, admitted the possibility that luminous AGNs 
dominated cosmic reionization. There were even widely 
discussed claims (Haiman \& Loeb 1999) that \chandra\ might detect 
$\approx 100$ quasars at $z\simgt 5$ in a single deep-field observation! 

\chandra\ observations allow the effective selection of AGNs, both
obscured and unobscured, that are up to $\approx 100$ times less 
luminous than those from wide-field optical surveys. 
These AGNs are $\simgt 500$ times more numerous. 
As a result, the AGN number counts from the deepest 
\chandra\ surveys have reached $\approx 7200$~deg$^{-2}$ 
(e.g., Bauer et~al.\ 2004; versus 
$\approx 13$~deg$^{-2}$ from the Sloan Digital Sky Survey
and $\approx 800$~deg$^{-2}$ from the deepest \rosat\ surveys). 
At a basic level, this is the key demographic ``discovery space'' 
that was opened by \chandra\ surveys.  

The moderate-luminosity AGNs discovered in the \chandra\ 
surveys are not measured to evolve in the same manner as luminous
quasars, indicating that AGN evolution is luminosity dependent
(e.g., Hasinger et~al.\ 2005; La~Franca et~al.\ 2005; 
Silverman et~al.\ 2008b; Yencho et~al.\ 2009; Aird et~al.\ 2010). 
Lower luminosity AGNs are found to 
peak in comoving number density at later cosmic times; 
this general behavior is sometimes referred to as ``cosmic downsizing''
or ``anti-hierarchical growth''. The details of this behavior 
are still somewhat uncertain owing to multiple thorny observational 
(e.g., detection incompleteness, source identification, 
follow-up incompleteness, \hbox{X-ray} spectral complexity) and 
statistical issues. Thus, the workers in this field often have 
strong, inconsistent opinions! However, the general 
consensus is that total SMBH power production peaks at significantly 
lower redshifts (\hbox{$z\approx 1$--1.5}) than expected based upon
evolution studies solely of luminous quasars (\hbox{$z\approx 2$--3}). 
At high redshift, the demographic constraints now show that there is 
indeed a decline in the space density of \hbox{X-ray} 
detected AGNs at $z\simgt 3$ (e.g., Fontanot et~al.\ 2007; 
Silverman et~al.\ 2008b; Brusa et~al.\ 2009a). 
This decline has a roughly exponential form, similar to what is 
found for optically selected quasars. Luminous AGNs are unlikely 
to have dominated cosmic reionization, leaving stars as the most 
likely agents.

\begin{figure}[t!]
\centerline{\includegraphics[width=.45\textwidth,angle=0]{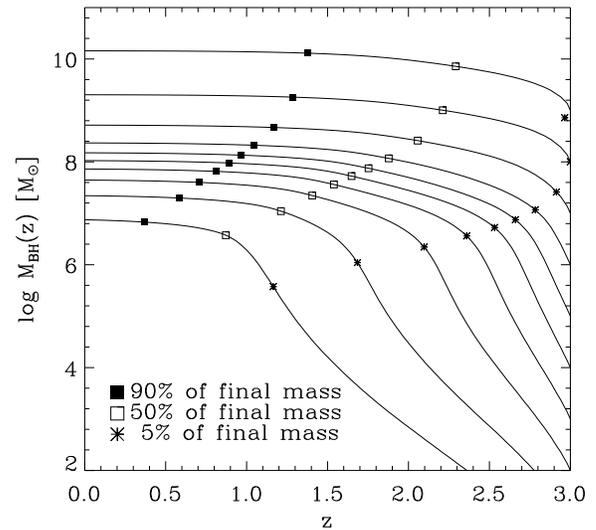}}
\caption{The average growth history of SMBHs as computed by 
Marconi et~al.\ (2006) using \hbox{X-ray} AGN luminosity functions. 
The symbols along each curve indicate the points where a 
SMBH reaches a given fraction of its final mass. Note that
more massive SMBHs grew at earlier cosmic times. SMBHs
which are now more massive than $\approx 10^8$~M$_\odot$ gained 
most of their final mass by $z\approx 1.5$, while lower mass 
black holes still grew substantially at lower redshifts.}\label{fig2}
\end{figure}

The luminosity functions delivered by the \hbox{X-ray} AGN demographers have
been used with versions of the elegant So\l tan (1982) argument to 
predict the masses of remnant SMBHs in galactic centers as well as 
the typical growth histories of SMBHs of various masses
(e.g., Marconi et~al.\ 2004, 2006; Merloni \& Heinz 2008; Shankar et~al.\ 2009). 
The most robust points generally emerging from this elaborate 
work are that standard radiatively efficient accretion can 
plausibly drive most SMBH growth, and that more massive SMBHs generally
grew earlier in cosmic time (e.g., see Fig.~2). 
Significant uncertainties still remain, however, in 
the luminosity functions themselves, 
the local SMBH mass function, 
bolometric corrections, 
Eddington ratios, and 
the efficiency of SMBH accretion. 
Together these limit the strength of some of 
the constraints that can be derived from So\l tan-type arguments. 

What has been the relative production of cosmic power by SMBHs vs.\,stars? 
Shortly before the \chandra\ launch, it was claimed that SMBHs may have 
supplied up to 50\% of the Universe's total energy output since the formation 
of galaxies (Fabian \& Iwasawa 1999). The \chandra\ AGN demographic results,  
however, now show that SMBH accretion has likely only supplied about 
\hbox{5--10\%} of this energy output; the remaining majority comes from 
nuclear fusion in stars. We appear to live in a remarkably economical 
\hbox{X-ray} universe, in that the observed cosmic \hbox{X-ray} 
background (CXRB) is produced with almost the least cosmic effort possible. It 
is not dominated by luminous obscured quasars thundering out huge amounts 
of power at \hbox{$z\approx 2$--4}, but rather by moderate-luminosity, 
obscured AGNs at \hbox{$z\approx 0.5$--2}. 

The work of the demographers is not finished. There is strong evidence 
that a large population of intrinsically luminous 
but heavily obscured (\hbox{$N_{\rm H}\simgt 3\times 10^{23}$~cm$^{-2}$}) 
AGNs, comprising a significant fraction of cosmic SMBH growth, is still
not detected in the \chandra\ surveys. This is not surprising, given expectations 
from the low-redshift universe. For example, many local Compton-thick 
(\hbox{$N_{\rm H}\simgt 1.5\times 10^{24}$~cm$^{-2}$}) 
AGNs that are intrinsically luminous
(e.g., NGC~1068, NGC~6240, Mrk~231)
would remain undetected even in the \chandra\ Deep Fields
if placed at \hbox{$z\simgt 0.5$--3}. 
Direct evidence for missed distant AGNs comes in several forms. For example, 
stacking analyses show that only \hbox{$\approx 50$--70\%} of the 
\hbox{6--8~keV} CXRB is resolved even in the deepest \hbox{X-ray} 
observations; the corresponding undetected \hbox{X-ray} source population 
plausibly has a sky density of \hbox{$\simgt 2000$--3000~deg$^{-2}$} with 
\hbox{$N_{\rm H}\approx 10^{23}$--$10^{24}$~cm$^{-2}$} at \hbox{$z\approx 0.5$--1.5} 
(e.g., Worsley et~al.\ 2005). 
Many compelling \hbox{X-ray} undetected AGN candidates 
have been found within the deepest \chandra\ observations via \spitzer\ surveys, 
radio surveys, optical-to-mid-infrared spectroscopy, and optical-variability 
studies (e.g., Alonso-Herrero et~al.\ 2006; Daddi et~al.\ 2007; 
Donley et~al.\ 2007, 2008; Alexander et~al.\ 2008a; Fiore et~al.\ 2008, 2009;
Treister et~al.\ 2009). These objects now require better characterization, 
at \hbox{X-ray} and other wavelengths, so that the contribution from SMBH 
accretion to their total luminosities can be determined reliably.


\section{Physics}

Extragalactic \chandra\ surveys have also provided insights into the 
processes shaping the observed \hbox{X-ray} emission from AGNs, ranging 
from accretion-disk (down to light minutes) to ``torus'' 
(\hbox{0.1--100}~light years) physical scales. They have served as
an essential complement to detailed \hbox{X-ray} studies of bright and 
usually nearby AGNs, often by providing powerful statistical constraints
upon the basic emission properties of moderate-luminosity, typical 
AGNs in the distant universe. 

When combined with multiwavelength AGN samples, 
\chandra\ surveys have been important in tightening empirical 
constraints upon the universality of \hbox{X-ray} emission from SMBH 
accretion disks and their so-called coronae (e.g., Mushotzky 2004; 
Brandt \& Hasinger 2005; Gibson et~al.\ 2008). This central dogma of 
universal \hbox{X-ray} emission (cf. Avni \& Tananbaum 1986), still 
on embarrassingly shaky ground from an {\it ab initio\/} physics 
point-of-view, underlies the utility of all \chandra\ surveys for 
finding AGNs throughout the Universe. 

The broad coverage of the luminosity-redshift plane provided by AGN 
samples in \chandra\ extragalactic surveys has allowed substantially
improved constraints to be set upon \hbox{X-ray}-to-optical/UV 
spectral energy distributions (SEDs; e.g., Steffen et~al.\ 2006; 
Just et~al.\ 2007; Kelly et~al.\ 2007, 2008; Gibson et~al.\ 2008; 
Green et~al.\ 2009; Young et~al.\ 2010). This is the spectral region 
where the direct accretion emission is dominant for relatively unobscured 
AGNs, and \hbox{X-ray}-to-optical/UV SED studies thus probe 
the inner \hbox{$\approx 100$--1000} gravitational radii
(e.g., the relative amounts of power emitted by the corona vs.\,the 
underlying disk). While there are still some 
discrepancies among published results (e.g., where fitted 
parameters from different samples disagree by much more than is allowed by their 
statistical uncertainties), some general points of consensus have emerged. 
First, there is a clear luminosity dependence of \hbox{X-ray}-to-optical/UV
SEDs for the majority population of radio-quiet AGNs, such 
that the ratio of \hbox{X-ray} vs.\,optical/UV emission declines with
rising optical/UV luminosity (e.g., see Fig.~3). This result, 
initially found in the 1980's with limited samples 
(e.g., Avni \& Tananbaum 1986), has now been established 
to hold out to \hbox{$z\approx 4$--6} and over a range of $\approx 100,000$ 
in luminosity. The form of the luminosity dependence is likely nonlinear, 
being stronger at high luminosities than low luminosities. Further work
to constrain this nonlinearity is required, as are {\it ab initio\/} 
physics-based calculations capable of predicting the luminosity dependence 
of \hbox{X-ray}-to-optical/UV SEDs (see, e.g., Noble \& Krolik 2009 and
references therein). 

The majority of current studies indicate that, after controlling for the 
luminosity dependence of \hbox{X-ray}-to-optical/UV SEDs, there is no
remaining detectable redshift dependence. For example, Steffen et~al.\ (2006)
and Just et~al.\ (2007) show that, at a fixed luminosity, the ratio 
of \hbox{X-ray}-to-optical/UV luminosity is constrained to change 
with redshift by $<30$\% out to \hbox{$z=5$--6}. It appears that, in spite 
of the large demographic changes in the AGN population over cosmic time, 
the individual AGN unit is remarkably stable on the scale of the inner 
accretion disk. 

\begin{figure}[t!]
\centerline{\includegraphics[width=.47\textwidth,angle=0]{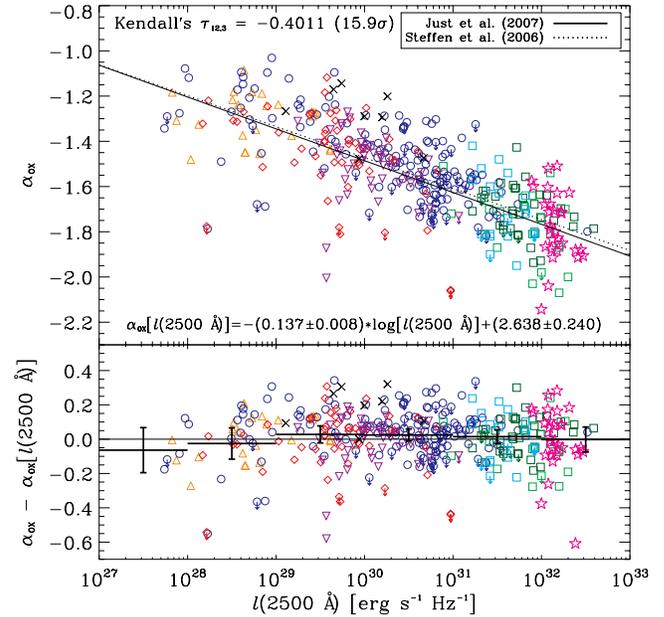}}
\vspace*{0.1 in}
\caption{One recent example showing the correlation between 
X-ray-to-optical/UV flux ratio, 
$\alpha_{\rm ox}=0.3838 \log(L_{\rm 2~keV}/L_{2500~\mathring{\rm{A}}})$, 
and the rest-frame 2500~\AA\ monochromatic luminosity for
radio-quiet AGNs; large negative values of \aox\ correspond to 
relatively weak \hbox{X-ray} emission. The different plotted symbols denote
the AGN samples utilized in the correlation analyses, ranging
from local Seyfert galaxies to the most-luminous quasars in the
Universe (the small number of downward-pointing arrows
denote \hbox{X-ray} upper limits). The 
\aox-$L_{2500~\mathring{\rm{A}}}$ relations from Steffen et~al.\ (2006) 
and Just et~al.\ (2007) are shown as dotted and solid lines, respectively, 
and the functional form of the dotted line is given at the bottom of
the top panel. The bottom panel shows residuals about the dotted 
line. The overlaid black error bars show, in $L_{2500~\mathring{\rm{A}}}$
bins, the mean of the residuals and the $3\sigma$ standard deviation 
of the mean. Adapted from Steffen et~al.\ (2006) and Just et~al.\ (2007), 
where details of the samples and fitting analyses are provided.}\label{fig3}
\end{figure}

Obscuration-based unification models have also been refined using
the large AGN samples from \chandra\ extragalactic surveys
(e.g., Ueda et~al.\ 2003; Barger et~al.\ 2005; 
La~Franca et~al.\ 2005; Treister \& Urry 2006; Hasinger 2008). 
Here again the broad coverage of the luminosity-redshift plane has
been essential, allowing obscuration dependences upon luminosity and
redshift to be constrained in much greater detail than was previously
possible. 
The improved data confirm longstanding expectations
(e.g., Lawrence \& Elvis 1982; Lawrence 1991) that the fraction of 
obscured AGNs drops with increasing luminosity; i.e., the covering factor 
of the torus is luminosity dependent, perhaps because more luminous AGNs
can evacuate their environments better. The obscured AGN fraction drops
in a roughly linear manner as a function of logarithmic 
\hbox{2--10~keV} luminosity, falling from 
$\approx 80$\% at $10^{42}$~erg~s$^{-1}$ to 
$\approx 20$\% at $10^{45}$~erg~s$^{-1}$.
Of course, the exact numerical values for these fractions depend upon 
how obscured AGNs are defined (\hbox{X-ray}, optical, and SED-based 
classification schemes do not consistently agree, especially at
low luminosities) and still have non-negligible systematic 
uncertainties owing to spectral complexity and missed AGNs. 

After controlling for luminosity effects, the fraction of obscured 
AGNs is found to rise with redshift as $(1+z)^{0.3-0.7}$
(e.g., La~Franca et~al.\ 2005; Treister \& Urry 2006; Hasinger 2008). 
This behavior appears to hold at least 
up to $z\approx 2$ where uncertainties become large
(systematic uncertainties, as mentioned above for the
luminosity dependence of the obscured fraction, are also 
relevant here). The processes ultimately shaping the
torus thus appear to evolve over cosmic time, in notable contrast 
to what is found for the inner accretion disk. The increase in
the covering factor of the torus with redshift is plausibly driven 
by the greater availability of gas and dust in galaxies at 
earlier cosmic epochs. 



\section{Ecology}

Since the launch of \chandra\, it has become well established that AGNs
play a role in the evolution of galaxies. The finding that many
massive galaxies in the local universe host a SMBH with a mass broadly
proportional to that of the galaxy spheroid hints at concordant
SMBH-spheroid growth (e.g., Tremaine et~al.\ 2002; H{\"a}ring \& Rix
2004), suggesting a close connection between AGN activity and star
formation. The optical-to-near-infrared emission from most of the distant 
AGNs detected in AGN surveys prior to \chandra\ was dominated by the active
nucleus, restricting the constraints that could be set upon 
host galaxies. Since the optical-to-near-infrared emission from a large
fraction of the \chandra-selected AGNs is dominated by starlight, it 
is now possible to measure directly the host-galaxy properties 
(e.g., morphology, color, luminosity, and stellar mass). By 
combining the \hbox{X-ray} data with infrared-to-radio observations, 
the relative power from AGN vs.\,star-formation activity can also 
be assessed.

High-resolution \hst\ imaging of $z\simlt1.5$ \hbox{X-ray} selected
AGNs in deep \chandra\ (and \xmm) surveys has shown that their host 
galaxies often have concentrated optical-light profiles, consistent with
expectations for galaxy spheroids (e.g., Grogin et~al.\ 2005; Pierce
et~al.\ 2007; Gabor et~al.\ 2009; Georgakakis et~al.\ 2009);
\hbox{$\approx$~40--50\%} appear to be early-type galaxies,
\hbox{$\approx$~20--30\%} appear to be late-type galaxies, and the rest are
peculiar or point-like systems. AGN host galaxies are also optically
luminous, indicating that they are massive 
[\hbox{$M_{\rm *}\approx$~(0.3--3)~$\times10^{11}$~$M_{\odot}$}; 
e.g., Babi{\'c} et~al.\ 2007; Alonso-Herrero et~al.\ 2008; 
Brusa et~al.\ 2009b]. First-order constraints therefore suggest that the 
SMBHs are comparatively massive and slow growing (typically 
$M_{\rm BH}\approx10^{8}$~$M_{\odot}$ and 
$L_{\rm Bol}/L_{\rm Edd}\approx10^{-2}$; e.g., Babi{\'c} et~al.\ 2007; Ballo
et~al.\ 2007; Alonso-Herrero et~al.\ 2008; Hickox et~al.\ 2009),
implying that they accreted the bulk of their mass at
$z\simgt 1.5$. Current constraints on the host-galaxy properties and
SMBH masses of $z\simgt 1.5$ AGNs are, however, poor due to the faintness 
(at optical-to-near-infrared wavelengths) of the majority of the population, 
and deeper imaging and spectroscopy are required for significant results
(see, e.g., Alexander et~al.\ 2008b, Brusa et~al.\ 2009b, and
Yamada et~al.\ 2009 for some constraints). Small rapidly growing SMBHs 
($M_{\rm BH}\simlt10^7$~$M_{\odot}$; $L_{\rm Bol}/L_{\rm Edd}>10^{-2}$) 
at $z<1$ are detected in deep X-ray surveys but appear to be 
comparatively rare (e.g., Ballo et~al.\ 2007; Shi et~al.\ 2008).

Similar to the normal-galaxy population at $z\simlt 1.5$, \hbox{X-ray} 
AGN host galaxies have a broad range of optical colors. However, while the
optical-color distribution of normal galaxies is clearly bimodal, with
a ``red sequence'' and ``blue cloud'', AGNs preferentially reside in
the red sequence, the top of the blue cloud, and the ``green valley'' 
in between (e.g., Nandra et~al.\ 2007; Silverman et~al.\
2008a; Hickox et~al.\ 2009). The green valley is the expected location
for galaxies transitioning between the blue cloud and the red sequence
due to the quenching of star formation (predicted by most 
galaxy formation models to be caused by large-scale outflows); 
however, bulge-dominated systems rejuvenated by the 
accretion of fresh gas from their environments could also lie in the
green valley (e.g.,\ Hasinger 2008). Sensitive spectroscopic 
observations could distinguish between these scenarios
by revealing the presence/absence of outflow signatures and 
cold accreted gas. 

The AGN host galaxies show no strong asymmetry when 
compared to non-AGN systems, indicating that they reside in 
relatively undisturbed systems. Contrary to some early
expectations, there also does not appear to be a connection between
recent strong galaxy mergers and moderate-luminosity AGN activity,
suggesting that SMBH growth is typically initiated by secular
host-galaxy processes and/or galaxy interactions (e.g., Grogin et~al.\
2005; Pierce et~al.\ 2007; Gabor et~al.\ 2009; Georgakakis
et~al.\ 2009). These results contrast with those found for rare, 
optically luminous quasars (\hbox{$\approx 100$--1000} times more 
luminous than the typical AGNs in \chandra\ blank-field surveys), 
which often appear to be associated with galaxy
major mergers (e.g., Canalizo \& Stockton 2001). These 
differences imply a change in the catalyst that drives the
fueling of luminous quasars and moderate-luminosity AGNs, as 
predicted by some models (e.g., Hopkins \& Hernquist 2009).
However, the fraction of $z\simlt 1$ galaxies hosting \hbox{X-ray} AGN 
activity appears to be enhanced in redshift filaments (slightly 
overdense regions) when compared to field-galaxy 
regions, suggesting that large-scale environment may help
drive SMBH growth (e.g., Gilli et~al.\ 2003; Silverman et~al.\ 2008a;
but see Georgakakis et~al.\ 2007 for potential host-galaxy mass
dependence). Differences in the AGN fraction between field galaxies
and galaxies in distant \hbox{(proto-)clusters} are also significant
and show that the bulk of SMBH growth in the densest regions occurred at
$z\gg 1$ (e.g., Lehmer et~al.\ 2009; Martini et~al.\ 2009).

\begin{figure}[t!]
\centerline{\includegraphics[width=.5\textwidth,angle=0]{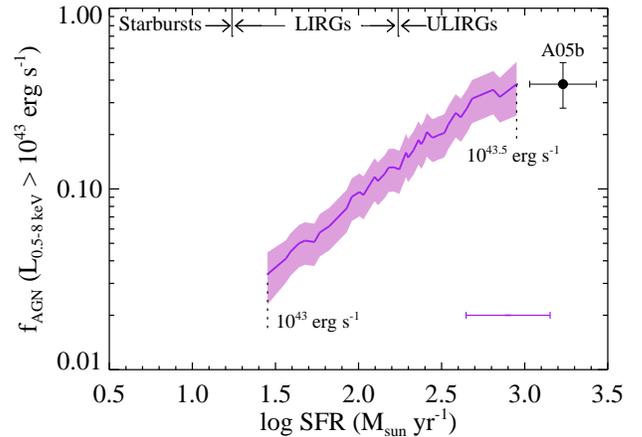}}
\caption{The dependence of the AGN fraction upon SFR for AGNs 
with \hbox{0.5--8~keV} luminosities above $10^{43}$~erg~s$^{-1}$; 
the dark-purple curve shows the best-estimated fraction, while the
light-purple region indicates the estimated uncertainty.  
The approximate AGN fraction for \hbox{$z\approx 2-3$} submm galaxies, 
from Alexander et~al.\ (2005b), is shown with the black data point. 
Approximate SFR ranges for starburst galaxies, 
luminous infrared galaxies (LIRGs), and 
ultraluminous infrared galaxies (ULIRGs) are shown along the
top. A ``sliding bin'' with a minimum of 10 AGNs was used to
construct this plot; the mean width of this bin is shown in
the lower right-hand corner. The minimum AGN \hbox{0.5--8~keV} 
luminosity sampled at the minimum and maximum SFR values 
is also indicated. From Rafferty et~al.\ (2010).}\label{fig4}
\end{figure}

The star-formation and SMBH-accretion histories broadly track each
other at least out to $z\approx 2$ (with an overall offset of a factor
of $\approx 5000$), as expected if the volume-averaged growth of
galaxies and SMBHs was concordant (e.g., Merloni et~al.\ 2004;
Silverman et~al.\ 2008b). Star formation in galaxies also
``downsizes'' in a qualitatively similar manner to what is seen for
AGNs (e.g., Damen et~al.\ 2009 and references therein). The majority of
individual \hbox{X-ray} selected AGNs have star-formation signatures
with implied star-formation rates (SFRs) of
$\approx$~1--1000~$M_{\odot}$~yr$^{-1}$ (e.g., Alexander et~al.\
2005ab; Pope et~al.\ 2008; Murphy et~al.\ 2009; Silverman et~al.\
2009), although the SFR vs.\,SMBH-accretion ratios for individual AGNs
can vary by several orders of magnitude. Using 70~$\mu$m \spitzer\
data, Rafferty et~al.\ (2010) have studied the \hbox{X-ray} AGN
fraction as a function of dust-obscured SFR for systems at
\hbox{$z\approx 0.2$--2.5} with 
\hbox{$L_{\rm IR}\approx10^{10}$--$10^{12}$~$L_{\odot}$}. They find that the
fraction of galaxies hosting \hbox{X-ray} moderate-to-luminous AGN
activity increases as a function of SFR, with an
\hbox{$\approx$~3--40\%} AGN fraction for SFRs of
\hbox{$\approx$~30--1000~$M_{\odot}$~yr$^{-1}$} (see Fig.~4), showing
directly that the duty cycle of moderate-luminosity AGN activity relates 
to the SFR of the host galaxy. The average AGN vs.\,star-formation
luminosity ratios for \hbox{X-ray} AGNs are found to be
broadly consistent with those expected from the local SMBH-spheroid mass
relationships, indicating a close connection between AGN activity and
star formation across a broad range of SFR. However, it is currently 
unclear whether the AGN-star formation connection is caused by regulatory 
feedback due to outflows (as adopted by some galaxy-evolution models) or 
some other process.


\section{Some Unresolved Questions and Future Prospects}

This concise review has provided a sampling of some of the significant
discoveries obtained by \chandra\ on the growth of SMBHs over cosmic
time. However, many important questions remain unanswered. Below we 
outline several of these along with prospects for future progress. 

{\bf Demography:} 
How many obscured AGNs are missed even in the deepest \hbox{X-ray} 
surveys, and what is their contribution to the growth of SMBHs? 
The current multiwavelength investigations have made great
advances in identifying \hbox{X-ray} undetected obscured AGNs, 
but all suffer from significant uncertainties (e.g., potential AGN
misidentifications, poorly constrained intrinsic AGN luminosities, 
small numbers of reliable identifications). Ultradeep \chandra\ and 
\xmm\ exposures, such as the upcoming 4~Ms \chandra\ Deep Field-South, 
will help to provide improved AGN characterization. Future sensitive
\hbox{$\approx 10$--200~keV} observations 
(e.g., with \nustar, \astroh, \ixo, and \exist), 
particularly when allied with improved data from 
multiwavelength facilities
(e.g., ELTs, \jwst, \herschel), will significantly extend the 
current census of SMBH growth in the most obscured systems. 

How do moderate-luminosity ($L_{\rm X}\approx10^{43}$~erg~s$^{-1}$) 
AGNs evolve over the important redshift interval of 
\hbox{$z\approx 2$--6} and beyond? 
Existing deep \hbox{X-ray} surveys already have the ability to 
detect high-redshift moderate-luminosity AGNs, provided their level 
of obscuration is not too strong, but it is often challenging 
to obtain accurate spectroscopic and/or photometric redshifts 
for these optically faint \hbox{X-ray} sources. Significant advances
in redshift determination can be made, e.g., with ultradeep 
(i.e., $>8$~hr) optical spectroscopy using the largest ground-based 
telescopes and with future large-area \hbox{X-ray}-to-millimeter 
observatories (e.g., \ixo, ELTs, \jwst, ALMA).
Larger \hbox{X-ray} survey areas at sensitive flux levels
(e.g., from \ixo, \erosita, and \wfxt\ observations) will also be 
essential for setting statistically powerful evolution constraints 
at the highest redshifts \hbox{($z\approx 4$--10)}. 

{\bf Physics:} 
Are there significant exceptions to the rule of universal \hbox{X-ray} 
emission from luminous AGNs?
Most of the \chandra\ AGN survey results are ultimately built 
upon the idea that strong underlying
\hbox{X-ray} emission is universally present. 
However, there are a small number of apparent 
\hbox{X-ray} weak exceptions to this 
rule that may be indicative of broader problems 
(e.g., Leighly et~al.\ 2007; Gibson et~al.\ 2008; and references therein). 
Surveys for further exceptions are important so that any foundational
cracks may be identified and patched. These surveys may
also lead to insights about accretion disks and their 
coronae. Strange objects, which persist in showing a type of spectrum 
entirely out of keeping with their luminosity, may ultimately teach us 
more than a host which radiates according to rule (cf. Eddington 1922)!

What is the nature of the luminosity dependence of the 
\hbox{X-ray}-to-optical/UV SEDs of AGNs? 
This (likely nonlinear) luminosity dependence still needs to be determined 
more reliably, since the current measurements of it quantitatively disagree
and thus cannot effectively guide the development of 
physical disk-corona models. A key aspect of future work must
be the reduction and realistic quantification of systematic errors 
including AGN misclassification, detection-fraction 
effects, absorption effects, host-galaxy light contamination, 
AGN variability, and luminosity dispersion. It is also critical to 
investigate further what practicable observables of AGN SEDs in 
the \hbox{X-ray}-to-optical/UV bandpass provide the most insight 
into their accretion processes, the roles of SMBH mass and Eddington 
fraction, and possible residual dependences of 
\hbox{X-ray}-to-optical/UV SEDs upon redshift. 

{\bf Ecology:} 
What are the host-galaxy properties of typical AGNs at $z>1.5$?
Although much has been revealed about the hosts of $z<1.5$
\hbox{X-ray} selected AGNs, comparatively little is known about the
(potentially more rapidly growing) hosts 
of $z>1.5$ \hbox{X-ray} AGNs. Currently, the biggest hindrance to
addressing this question is the lack of rest-frame 
optical-to-near-infrared observations with the requisite 
combination of high sensitivity and angular resolution. 
This situation should significantly improve with \hst\ WFC3
and \jwst\ rest-frame optical-to-near-infrared observations
in the future. 

What is the physical meaning of the color-magnitude diagram results
for AGN host galaxies? 
It is currently unclear the extent to which 
the green-valley and red-sequence colors for 
typical \hbox{X-ray} AGNs at $z<1.5$ are due to 
the quenching of star formation, 
the rejuvenation of bulge-dominated systems,
dust extinction, 
biases in sample construction, or 
something else. 
Spatially resolved spectroscopy of individual sources can be used 
to search for the large-scale outflow signatures expected to
quench star formation (e.g., Nesvadba et~al.\ 2008; Alexander et~al.\
2010), and millimeter spectroscopy (e.g., with existing facilities or
ALMA in the future) can provide constraints on the presence of 
cold molecular gas. 

What are the effects of cosmic environment, from voids to
superclusters, on the growth of SMBHs? 
Given the different evolution of AGNs in \hbox{(proto-)clusters} from those 
detected in blank-field \hbox{X-ray} surveys (e.g., Lehmer et~al.\ 2009; 
Martini et~al.\ 2009), it is clear that 
environment must play some role in the growth of SMBHs. \hbox{X-ray}
surveys are required with sufficient areal coverage {\it and\/} sensitivity
to identify the AGNs that dominate the \hbox{X-ray} luminosity function
across the full range of redshifts {\it and\/} environments. This can be
accomplished with large investments of \chandra\ and \xmm\ time as well as
future facilities (e.g., \ixo, \wfxt). 






\begin{acknowledgments}
We thank all of our collaborators on Chandra extragalactic surveys for
educational interactions over the past decade. 
We thank M. Brusa and an anonymous referee for helpful feedback on
this paper. 
We gratefully acknowledge support from 
Chandra Award SP8-9003A (WNB), 
NASA ADP grant NNX10AC99G (WNB), 
the Royal Society (DMA), and 
a Philip Leverhulme Prize (DMA). 
\end{acknowledgments}





\end{article}








\end{document}